 \documentclass[final,2p,times]{elsarticle}
\usepackage{amsfonts}
\usepackage{amsmath}
\usepackage{amssymb}
\usepackage{mathrsfs}
\usepackage[english,activeacute]{babel}
\usepackage[latin1]{inputenc}
\usepackage{graphicx}
\usepackage{subfigure}
\usepackage{float}

 \usepackage{color}

\begin{document}

\title{Existence of periodic orbits in nonlinear oscillators of Emden-Fowler form}

\author{Stefan C. Mancas}
\ead{mancass@erau.edu}
  \address[label1]{Department of Mathematics, Embry--Riddle Aeronautical University\\ Daytona Beach, FL 32114-3900, USA}

\author{Haret C. Rosu\corref{cor1}}
 \ead{hcr@ipicyt.edu.mx}
  \address[label2]{IPICyT, Instituto Potosino de Investigacion Cientifica y Tecnologica,\\
Camino a la presa San Jos\'e 2055, Col. Lomas 4a Secci\'on, 78216 San Luis Potos\'{\i}, S.L.P., Mexico}

\cortext[cor1]{Corresponding author}
%

%\pacs{02.30.Ik, 05.45.-a}
%Integrable systems, NL dynamics and Chaos

\begin{abstract}

\noindent The nonlinear pseudo-oscillator recently tackled by Gadella and Lara %$yy''+1=0$
is mapped to an Emden-Fowler (EF) equation that  is written as an autonomous two-dimensional ODE system for which we provide the phase-space analysis and the parametric solution. Through an invariant transformation we find periodic solutions to a certain class of EF equations that pass an integrability condition. We show that this condition is necessary to have periodic solutions and via the ODE analysis we also find the sufficient condition for  periodic orbits. EF equations that do not pass integrability conditions can be made integrable via an invariant transformation which also allows us to construct periodic solutions to them. Two other nonlinear equations, a zero-frequency Ermakov equation and a positive power Emden-Fowler equation are discussed in the same context.

\end{abstract}

\begin{center}
Phys. Lett. A (2016)
\end{center}

\maketitle

\noindent {\bf Highlights}\\

\begin{itemize}

\item An invariant transformation is used to find periodic solution of EF equations.
\item Phase plane study of the EF autonomous 2D ODE system is performed.
\item Three examples are presented from the standpoint of the phase plane analysis.

\end{itemize}

\newpage

%11111111111111111111111

\subsection*{1. Introduction}

Many nonlinear oscillators belonging to the classes of positive and negative power nonlinearity can be put in the following Emden-Fowler form \cite{Fow}
\begin{equation}\label{ef}
q_{_{YY}}-\alpha Y^{-\lambda-2}q^n=0~.
\end{equation}
For $\lambda=-2$ and written in self-adjoint form, see Eq.~(\ref{EFnormal}) below, Eq.~(\ref{ef}) is known as the Lane-Emden equation and emerged first in astrophysics as the equation for the Newtonian gravitational potential of a spherically symmetric polytropic gas \cite{Lane,Emden},
with the dependent variable related to the density of the self-gravitating gas and the independent variable as a dimensionless radius. The Lane-Emden equation is actually the Poisson equation in disguise for polytropic gases in convective equilibrium. The case with $n=3$ and the dependent variable taken as the thermodynamic ideal temperature was used by Eddington in his theory of the internal constitution of stars which led to the famous mass-luminosity relation for these cosmic objects \cite{Edd}. Moreover, the Thomas-Fermi model \cite{Thomas,Fermi} for the electrostatic field in the bulk of a heavy atom has the same Poisson background and is also expressed by the self-adjoint form of Eq.~(\ref{ef}) although for a non-integer $n$. On the other hand, for negative $n$ of the form $n=1-2m$, with $m$ a positive integer $\geq 2$, the self-adjoint Emden-Fowler equations can be associated to Ermakov parametric oscillators and their Reid generalization \cite{mr}, while for $m=1$, it gives a simple model of the path taken by an electron in an electron beam injected into a plasma tube \cite{Act-Sq}.

\medskip

An important problem for nonlinear differential equations is to examine the existence of periodic solutions around some critical point. It appears that for the subclass of Emden-Fowler equations defined by $\lambda=-2$ and any negative exponent $n$ this problem is still under some debate. For example, in a recent paper by Gadella and Lara (GL)~\cite{ref1}, it was argued that the particular case of (\ref{ef})
%......1
\begin{equation}\label{e0}
q q_{_{YY}}+1=0~,
\end{equation}
also known as the `pseudo-oscillator', has no periodic oscillatory  solutions despite previous claims in the literature, which were based on approximate solution methods of this equation. On the other hand, Van Gorder \cite{vG} showed that while smooth periodic solutions, i.e., with continuous derivatives, may not exist, non-smooth continuous periodic solutions (with the derivative not continuous at some points on the real axis) can still be constructed.
The GL paper served us as a motivation to study this problem in the more general Emden-Fowler formulation and the corresponding autonomous two-dimensional ODE  system. After a brief discussion of the general pseudo-oscillator solution in section 2, we introduce the integrable Emden-Fowler cases according to Rosenau \cite{ros} in section 3, where we show that there is a transformation of variables through which the pseudo-oscillator is made Rosenau integrable. In section 4, we present the phase-plane analysis of the ODE system equivalent to the Emden-Fowler equations. Three illustrative examples, including the pseudo-oscillator, are discussed from this standpoint in section 5. Our conclusions are presented in section 6, essentially stating that smooth continuous periodic solutions exist only for the positive class of single power nonlinearity when $n=2 \lambda+1>1$.
%The latter condition is not fulfilled by the pseudo-oscillator.

\subsection*{2. General solution of the pseudo-oscillator equation}

\begin{figure}[htb!]
 \centering
 \resizebox*{0.5\textwidth}{!}{\rotatebox{0}
{\includegraphics{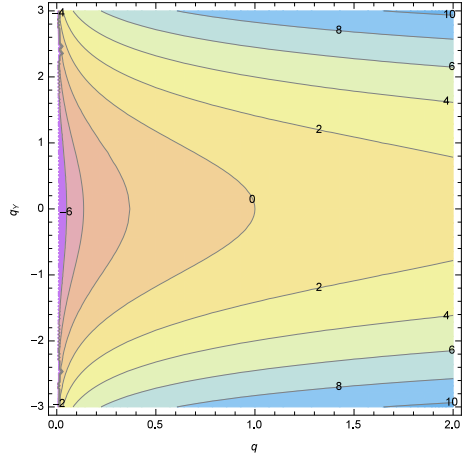}}}
 \caption{\label{figone} Hamiltonian curves with the corresponding values of ${\mathcal H}$ in the phase space $q,q_{_Y}$.}
\end{figure}

By multiplying by $q_{_Y}$ in (\ref{e0}) one can easily find the first integral of motion
%..........................2
\begin{equation}\label{pseudo}
(q_{_Y})^2+\ln q^2=\mathcal H~,
\end{equation}
where ${\mathcal H}$ denotes a Hamiltonian with logarithmic potential $V(q)=2 \ln q$, see Fig. \ref{figone}.

\begin{figure}[htb!]
 \centering
 \resizebox*{0.4\textwidth}{!}{\rotatebox{0}
{\includegraphics{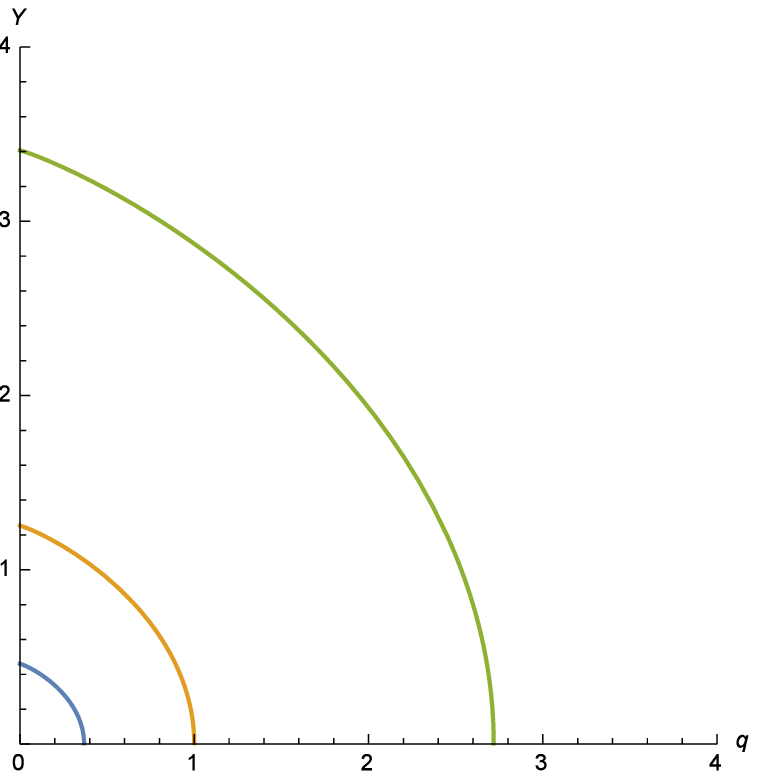}}}
 \resizebox*{0.5\textwidth}{!}{\rotatebox{0}
{\includegraphics{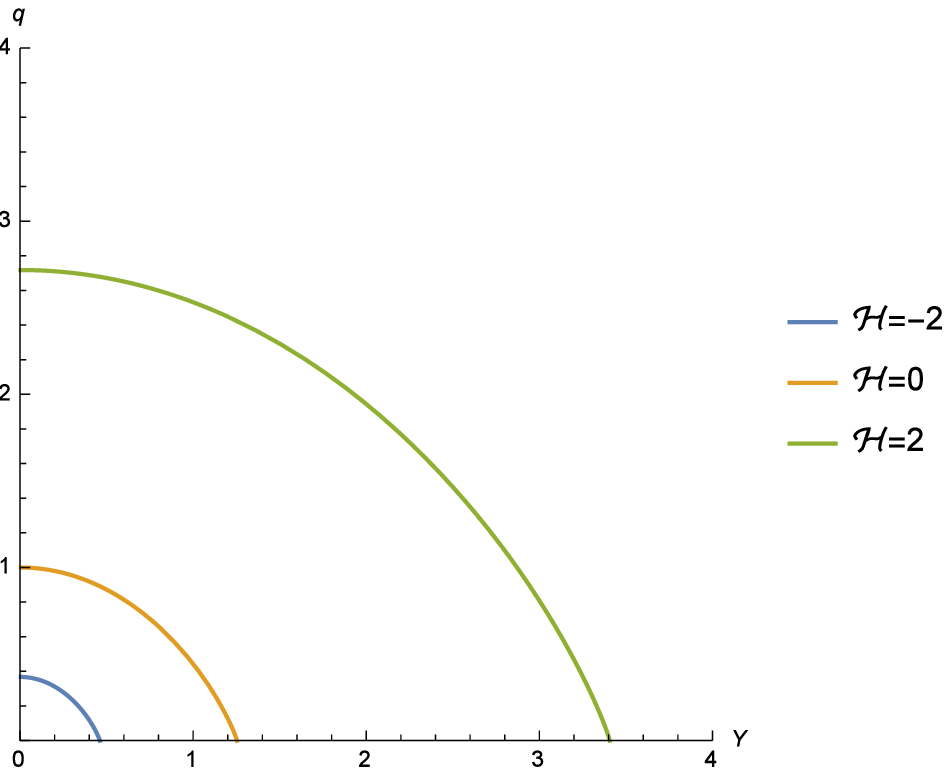}}}
 \caption{\label{figtwo} Solutions (\ref{pol}) and (\ref{gen}) to equation (\ref{e0}).}
\end{figure}

By a quadrature one can get the Polyanin solution \cite{pol}
%.........................3
\begin{equation}\label{pol}
Y-Y_0=\pm\int^q \frac{d z}{\sqrt{\mathcal H-2 \ln z}}=\mp A\,%\sqrt{\frac \pi 2}\exp\left(\frac c 2\right)
\mathrm{erf}\left(\sqrt{\frac{\mathcal H}{2}-\ln q}\right)~,
\end{equation}
where the amplitude is
$$
A=\sqrt{\frac \pi 2}\exp\left(\frac {\mathcal H} {2}\right)~.
$$
This solution can be identified with the solution given by Gadella and Lara in their Eq.~(7) if $Y_0=c_2$ and $\frac{\mathcal H}{2}=-c_1$, where $c_1$ and $c_2$ are the constants of Gadella and Lara. The general pseudo-oscillator solution is obtained by inverting \eqref{pol}, which gives
%...........................
\begin{equation}\label{gen}
q(Y)=\sqrt{\frac 2 \pi}A\exp\Bigg\{-\left[{\rm erf}^{-1}\left(\mp \frac{Y-Y_0}{A}\right)\right]^2\Bigg\}~,
\end{equation}
see Fig.~\ref{figtwo} for $Y_0=0$, and $\mathcal H=-2,0,2$.

Gadella and Lara claim that this solution is not oscillatory, or in more general terms that in the phase space there is {\em no closed orbit associated to a periodic solution surrounding at least a critical point}, which is a necessary fingerprint for periodic solutions.

\subsection*{3. The Emden-Fowler approach and an invariant transformation}

 The self-adjoint form of Eq.~(\ref{ef}) is obtained by using Kamke's substitutions \cite{Kamke}, $q(Y)=\eta(\xi)$ and $\xi=\frac{1}{Y}$,
 which lead to
 \begin{equation}\label{EFnormal}
\frac {d}{d \xi}(\xi^2 \eta\rq{})=\alpha\xi^{\lambda}\eta^{n}~,
 \end{equation}
 which becomes
 \begin{equation}\label{EFnormal2}
\xi \eta \rq{}\rq{}+2 \eta\rq{}=\alpha\xi^{\lambda-1}\eta^{n}~,
 \end{equation}
where $\rq{}=d/d \xi$.

Recall now that in 1984 Rosenau \cite{ros} was interested in the integration of the above equation for which he constructed integrals of motion, provided that two conditions are satisfied:  $n=2 \lambda+1$ or $n=\lambda-1$.
Therefore  EF equations of type
\begin{eqnarray}\label{lg}
\begin{array}{ll}
&q_{_{YY}}=\alpha Y^{-\lambda-2}q^{2 \lambda+1}\\
&q_{_{YY}}=\alpha Y^{-\lambda-2}q^{\lambda-1}\\
\end{array}
\end{eqnarray}
 have integrals of motion and are integrable by quadratures.

 In 1970s, Djukic \cite{Djuk} was concerned with finding integrals of motion for EF equations of the type
\begin{equation}\label{1000}
t \ddot x+2 \dot x+a t^\nu x^{2 \nu +3}=0
\end{equation}
which are identical to Eq.~(\ref{EFnormal2}) provided that $a=-\alpha, ~n=2(\lambda-1)+3=2 \lambda+1,~ \nu=\lambda-1$
and he found that an integral of motion for Eq.~(\ref{1000}) is

\begin{equation}\label{1001}
t^3 \dot x^2+t^2 x \dot x +\frac{a}{\nu+2}t^{\nu+2}x^{2(\nu+2)}=const.
\end{equation}
According to our identification of constants the integral of motion that corresponds to the first Rosenau integrability condition $n=2 \lambda+1$ is

 \begin{equation}\label{99}
\xi^3 \eta\rq{}^2+\xi ^2\eta \eta\rq{}-\frac{\alpha}{\lambda+1}\xi^{\lambda+1}\eta^{2(\lambda+1)}={\mathcal C},
 \end{equation}
 which in terms  of the original variables of Eq. (\ref{ef}) turns into
  \begin{equation}\label{100}
Y (q_{_Y})^2-q q_{_Y}-\frac{\alpha}{\lambda+1}\left(\frac{q^2}{Y}\right)^{\lambda+1}={\mathcal C}.
 \end{equation}

Since for the pseudo-oscillator we require $n=-1, \lambda=-2$ none of the integrability conditions found by Rosenau are satisfied. However, we present a transformation allowing us to circumvent this problem for which the second condition will be satisfied.

\medskip

Let us use

\begin{eqnarray}\label{lha}
\begin{array}{ll}
&q=\frac w s\\
&Y=\frac 1 s\\
\end{array}
\end{eqnarray}
that we call an invariant transformation, then the EF equation in~(\ref{ef}) can be written with a different power of the independent variable
%..........................
\begin{equation}\label{eq1}
w_{ss}=\alpha s^{\lambda -1-n}w^n.
\end{equation}
Now, letting $n=2\lambda+1$  we then obtain
\begin{equation}\label{lh}
w_{ss}=\alpha s^{-\lambda-2}w^{2 \lambda+1}.
\end{equation}
This equation is the same as (\ref{ef}) with the first Rosenau condition fulfilled, which is a fingerprint of the invariant transformation, and hence the self-adjoint form~(\ref{EFnormal}) is recovered for $n=2\lambda+1$, which is another feature of the invariant transformation.

   On the other hand, if we use the second Rosenau integrability condition $n=\lambda-1$, we obtain the `partner' equation
%..........................
\begin{equation}\label{lgh}
\tilde{w}_{ss}=\alpha \tilde{w}^{\lambda-1}~,
\end{equation}
which is the pseudo-oscillator when $\lambda=0$, and the linear oscillator when $\lambda=2$.
Moreover, now we can write the integral of motion for the case $n=\lambda-1$ by multiplying by $\tilde w_s$ and integrating to get
%..........................
\begin{equation}\label{1004}
\tilde {w}_s^2-\frac{2 \alpha}{\lambda}\tilde w^\lambda=\mathcal{C}
\end{equation}
which in terms of the original variables of Eq.~(\ref{ef}) becomes
\begin{equation}\label{1005}
(q-Yq_Y)^2-\frac{2 \alpha}{\lambda}\left(\frac{q}{Y}\right)^\lambda=\mathcal{C}.
\end{equation}

Thus, by varying  $\lambda$ in (\ref{lgh}) one can generate classes of EF equations of type (\ref{lh}) which, according to the analysis presented in the next section, will have periodic solutions for $\lambda>1$.

In addition, Rosenau integrability conditions  allows us  to write general parametric solutions of both Eqs.~(\ref{lg}) %(7)
as follows:

(i) The first equation of system~(\ref{lg}) %(7)
is  Eq.~(4) in section \S 2.3.1-2 of  Polyanin\rq{}s book

\begin{equation}\label{equ1}
y\rq{}\rq{}=Ax^{-\frac{n+3}{2}}y^{n}~,
\end{equation}

where $A=\alpha $ and $n=2 \lambda+1$.

%Depending on $\lambda$ there are two sets of parametric solutions.
\begin{eqnarray}\label{equ2}
\begin{array}{ll}
&Y(\tau)=a C_1^2\exp \Theta(\tau)\\
&q(\tau)=b C_1 \tau \exp{\frac{1}{2}\Theta (\tau)}\\
&\Theta(\tau)=\int{\frac{d \tau}{\sqrt{C_2+\psi(\tau)+\frac{\tau^2}{4}}}}\\
& \alpha=\left(\frac{a}{b^2}\right)^\lambda~.
\end{array}
\end{eqnarray}
Depending on $\lambda$ there are two sets of parametric solutions:

\noindent First, if $\lambda \ne -1$ we have
$\psi(\tau)=\frac{\tau^{2(\lambda+1)}}{\lambda+1}$, while if $\lambda=-1$ then $\psi(\tau)= 2 \ln |\tau|$.

Using system \eqref{equ2} then
\begin{equation}\label{xixi}
q^2=\frac{\tau^2}{\alpha^{\frac 1 \lambda}}Y~.
\end{equation}
(ii) The second equation of system~(\ref{lg}) 
is Eq.~(3) in section \S 2.3.1-2 of  Polyanin\rq{}s book

\begin{equation}\label{equ4}
y\rq{}\rq{}=Ax^{-(n+3)}y^{n}~,
\end{equation}
where $A=\alpha $ and $n=\lambda -1$.
So, depending on the sign of $\lambda$ there are also two sets of parametric solutions:

\noindent If $\lambda \ne 0$ we have
\begin{eqnarray}\label{equ5}
\begin{array}{ll}
&Y(\tau)=\frac{a C_1~^{\lambda-2}}{\Theta(\tau)}\\
&q(\tau)=\frac{b C_1^{\lambda}\tau}{\Theta(\tau)}\\
&\Theta(\tau)=C_2+\int{\frac{d \tau}{\sqrt{1\pm \tau^ \lambda}}}\\
& \alpha=a^\lambda b^{2-\lambda}\frac \lambda 2~,
\end{array}
\end{eqnarray}
while, if $\lambda=0$, we have
\begin{eqnarray}\label{equ6}
\begin{array}{ll}
&Y(\tau)=\frac{C_1}{\Theta(\tau)}\\
&q(\tau)=\frac{b \exp{\mp \tau^2}}{\Theta(\tau)}\\
&\Theta(\tau)=C_2+\int{\exp{\mp \tau^2} d \tau}\\
& \alpha=\pm 2b^2~.
\end{array}
\end{eqnarray}

\section*{4. An autonomous two-dimensional ODE system}
By transforming the general EF equation into an autonomous two-dimensional ODE system one can classify the solutions based on linear stability analysis.
This mapping can be achieved by using the transformations given by Jordan and Smith in \cite{js}
\begin{eqnarray}\label{l1}
\begin{array}{ll}
&X=\frac{\xi \eta'}{\eta}\label{v1}\\
&Y=\xi^{\lambda -1}\frac{\eta^{n}}{\eta'}~,\label{v2}
\end{array}
\end{eqnarray}
with  $\xi=e^t$  will turn (\ref{EFnormal}) into an autonomous two-dimensional ODE system
\begin{eqnarray}\label{l2}
\begin{array}{ll}
& \dot{X}=-X(1+X-\alpha Y)=M(X,Y)\\
&\dot{Y}=Y(1+\lambda +n X-\alpha Y)=N(X,Y)~,
\end{array}
\end{eqnarray}
where  $\dot {}\, =d/dt$ and with the four equilibrium points given by
$$
\left\{ (X_0,Y_0)=(0,0);  ~(X_1,Y_1)=(-1,0);~  (X_2,Y_2)=(0,\frac{\lambda+1}{\alpha}); ~(X_3,Y_3)=\left(-\frac{\lambda}{n-1}, \frac {\lambda-n+1}{\alpha(1-n)}\right) \right\}.
$$

Following standard methods of phase-plane analysis, we  use the linear approximation of the equilibrium points  to classify them. The Jacobian matrix of (\ref{l2}) is

\begin{eqnarray}\label{17}
J=\left[\begin{array}{cc}
\frac{\partial M}{\partial X}& \frac{\partial M}{\partial Y}\\
\frac{\partial N}{\partial X}& \frac{\partial N}{\partial Y}\\
\end{array}\right]=\left[\begin{array}{cc}
-1-2X+\alpha Y& \alpha X\\
nY& 1+\lambda +nX- 2 \alpha Y\\
\end{array}\right]
\end{eqnarray}
and the characteristic polynomial of the Jacobian matrix is
%........................eq. 15
\begin{equation} \label{charac1}
\theta^2-\delta_1 \theta+\delta_2=0~.
\end{equation}

The equilibrium points will be classified according to signs of the trace $\mathrm{Tr}(J)=\delta_1=\frac{\partial M}{\partial X}+ \frac{\partial N}{\partial Y}$, the determinant $\mathrm{Det}(J)=\delta_2=\frac{\partial M}{\partial X}\frac{\partial N}{\partial Y}-\frac{\partial M}{\partial Y}\frac{\partial N}{\partial X}$, and the discriminant $\Delta=\delta_1^2-4\delta_2$, all evaluated at $(X_i,Y_i)$.

\medskip

As we can see from the Table~\ref{tab1} the location in the phase space depends on the nonlinear coefficient $\alpha$ and the powers $\lambda, n$, while the type of fixed point (its classification) is given only by the powers $\lambda, n$.

In order to have purely periodic solutions a center is obtained when $\delta_1=0$ and $\delta_2>0$   which tells that the only  fixed point  that could be a center is $(X_3,Y_3)$. Therefore the  curve  $\delta_1=0$ is given exactly by the Rosenau first  integrability  condition $n=2\lambda+1$ which  is a {\em necessary} condition for periodic solutions. Using this condition we obtain  $\delta_2=\frac \lambda 2$ which provides the {\em sufficient} condition for periodicity, namely $\delta_2>0 \Rightarrow n>1$.

\begin{table}[htb!] %
\begin{center}
\begin{tabular}{|c|c|c|c|c|c|c|}
\hline Fixed Points & $\delta_1$ & $\delta_2$ & $\Delta$ & Type \\
\hline
\hline
$(X_0,Y_0)$ & $\lambda$ & $-(1+\lambda)$ & $(\lambda+2)^2$   & saddles, nodes (stable/unstable)\\
\hline
$(X_1,Y_1)$  & $2-n+\lambda $ & $1-n+\lambda$ & $(n-\lambda)^2$  &  saddles, nodes (stable/unstable) \\
\hline
$(X_2,Y_2)$ & -1 & $-\lambda(1+\lambda)$  & $(1+2 \lambda)^2$ & saddles, nodes (stable) \\
\hline
$(X_3,Y_3)$ & $\frac{1-n+2 \lambda}{-1+n}$ & $\frac{(-1+n-\lambda)\lambda}{-1+n}$ & $\frac{1+n[-2+n-4n\lambda+4\lambda(1+\lambda)]}{(-1+n)^2}$ & all\\
\hline
\end{tabular}
\end{center}
\caption{General equilibrium points of the autonomous two-dimensional ODE system (\ref{l2}).}
\label{tab1}
\end{table}

\section*{5. Examples}
We present now three cases of nonlinear ODEs for which the periodicity of solutions is characterized using the above phase-plane analysis. In all examples we use  $\alpha=-1$.

\medskip

%...............................
\noindent (i) {\em  Ermakov-type equation}.
For $\lambda=-2 \Rightarrow n=-3$, all of Eqs.~(\ref{ef}), (\ref{lg}), (\ref{eq1}), and (\ref{lh}) are Ermakov equations of zero frequency
%for which $\lambda=-2 \Rightarrow n=-3$
%........ eq. 16 .........
\begin{equation}\label{er}
q^3q_{_{YY}}+1=0~.
\end{equation}

There are two invariants that the latter equation possesses. One is the well-known Ermakov invariant
\begin{equation}\label{er3}
{\mathcal I}=\frac 1 2 [(q_{_Y})^2-q^{-2}]~,
\end{equation}
and the second is the ${\mathcal C}$ invariant, which by using (\ref{100}) and (\ref{er3}) is
%....... eq 19 .........
 \begin{equation}\label{er4}
{\mathcal C}= Y[(q_{_Y})^2-q^{-2}]-qq_{_Y}=2IY-qq_{_Y}~,
\end{equation}
and also shows that $(q^2)_{_Y}$ is a linear function of $Y$ and the invariant ${\mathcal C}$ can be determined from the ordinate intersection. By solving this equation and choosing an appropriate integration constant, we get the general solution of Eq.~(\ref{er}) as a function of ${\mathcal C}$
 \begin{equation}\label{er4a}
q(Y)=\sqrt{1-2 \mathcal C Y+(\mathcal C^2-1)Y^2}~.
\end{equation}
As a particular solution, if one chooses $\mathcal C=0$ we recover
 Pinney\rq{}s solution which comes from superposition formula \cite{Pin}
\begin{equation}\label{er2}
q(Y)=\sqrt{1-Y^2}~.
\end{equation}

Because the equation stays invariant under the transformation, another solution can be obtained from (\ref{er2}) to get
\begin{equation}\label{er20}
w(s)=\sqrt{s^2-1}~,
\end{equation}
which will solve
\begin{equation}\label{err}
w^3w_{_{ss}}+1=0~.
\end{equation}

Since this equation  is both of the type  (\ref{equ1}) or   (\ref{equ4}) for $n=-3$ for which $\lambda=-2$, then for appropriate choices of constants $C_1,C_2,a ,b$ we can use both sets of the parametric solutions  (\ref{equ2}) or  (\ref{equ5}) to obtain the general solution (\ref{er4a}), or a particular solution such as (\ref{er2}) or (\ref{err}).

\begin{figure}[htb!]
 \centering
\resizebox*{0.75\textwidth}{!}{\rotatebox{0}
{\includegraphics{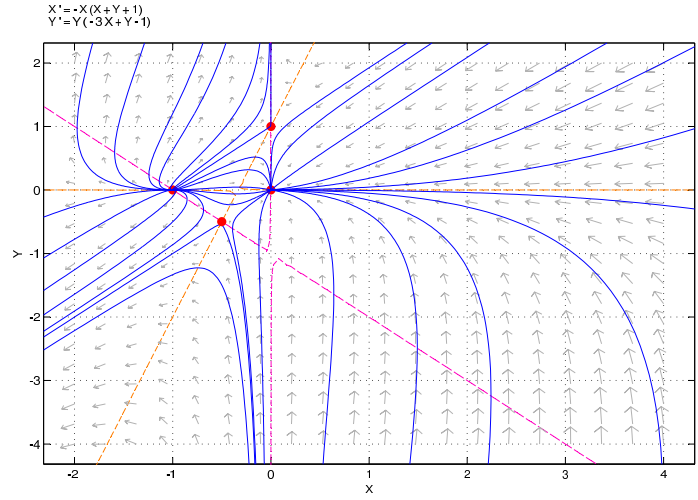}}}
 \caption{\label{figureone} Phase plane portrait for the Ermakov equation (\ref{er}).}
\end{figure}

Since for this case $(\delta_1,\delta_2)=(0,-1)$, then $(X_3,Y_3)=(-\frac 1 2,-\frac 1 2)$ becomes a saddle, no periodic solutions are allowed because $\lambda<0$,  see Fig.~\ref{figureone}.
%..........Figure 3 ..........

\medskip

%..................................................
\noindent (ii) {\em The pseudo-oscillator equation}.

For $\lambda=-2$, $n=-1$, Eq.~(\ref{ef}) is the pseudo-oscillator Eq.~(\ref{e0})
and the first Rosenau integrability condition is not satisfied. If we now use the invariant transformation (\ref{lha}) we obtain the EF with different powers
\begin{equation}\label{nee}
w_{ss}=\alpha s^{-2}w^{-1}
\end{equation}
which is actually of the type (\ref{equ4}) with $n=-1$. Therefore, now the second Rosenau integrability condition is satisfied and one can use the parametric system (\ref{equ6}) with $\lambda=0$ to generate solutions of the pseudo-oscillator equation, including solution (\ref{gen}).

\begin{figure}[htb!]
 \centering
\resizebox*{0.75\textwidth}{!}{\rotatebox{0}
{\includegraphics{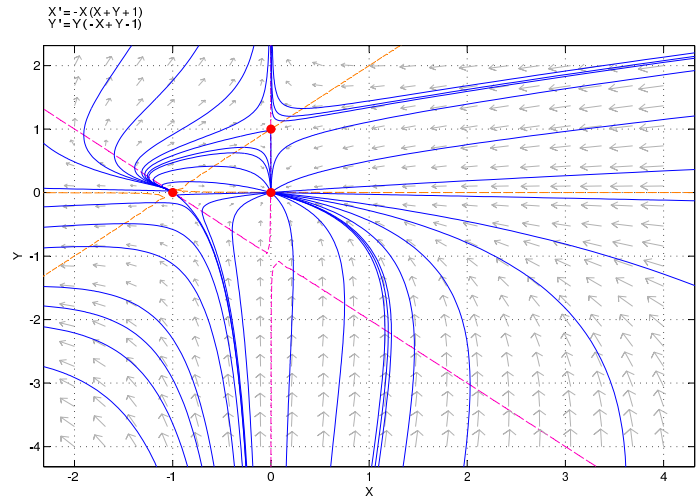}}}
 \caption{\label{figuretwo} Phase plane portrait for the pseudo-oscillator equation (\ref{e0}).}
\end{figure}

Since  $(\delta_1,\delta_2)=(1,0)$ then the two fixed points collide $(X_3,Y_3)=(X_1,Y_1)=(-1,0)$  becoming a degenerate
unstable node as seen in Fig.~\ref{figuretwo}.

%...... Figure 2 ...........

\medskip

%................................................
\noindent (iii) {\em A positive power EF equation}.
Let us choose $\lambda =\frac 1 2 \Rightarrow n=2$, which gives $(X_3,Y_3)=(-\frac 1 2, -\frac 1 2)$ a center, since for this case  $(\delta_1,\delta_2)=(0,\frac 1 4)$.

\begin{figure}[htb!]
 \centering
 \resizebox*{0.6\textwidth}{!}{\rotatebox{0}
{\includegraphics{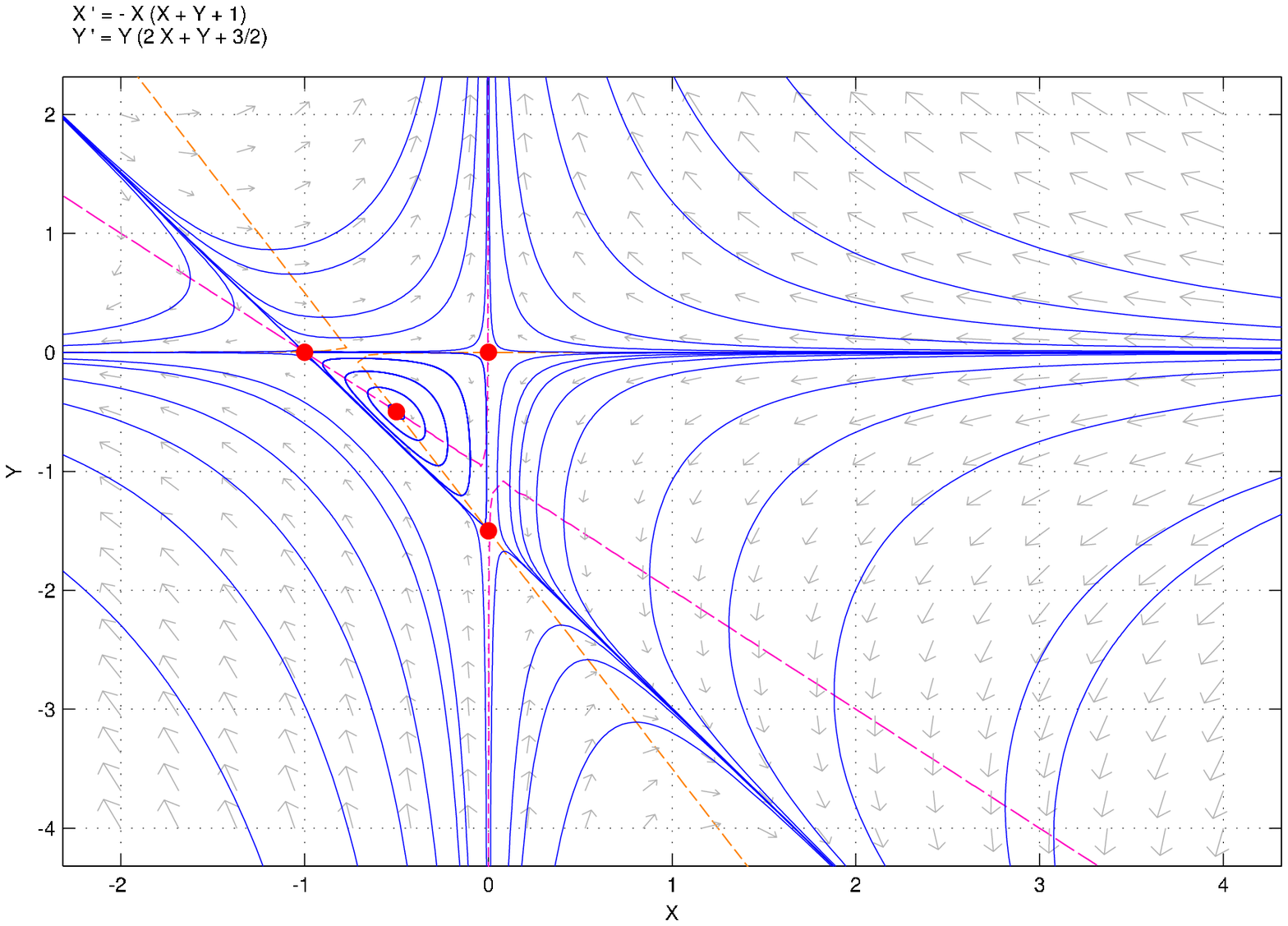}}}\\
 \resizebox*{0.5\textwidth}{!}{\rotatebox{0}
{\includegraphics{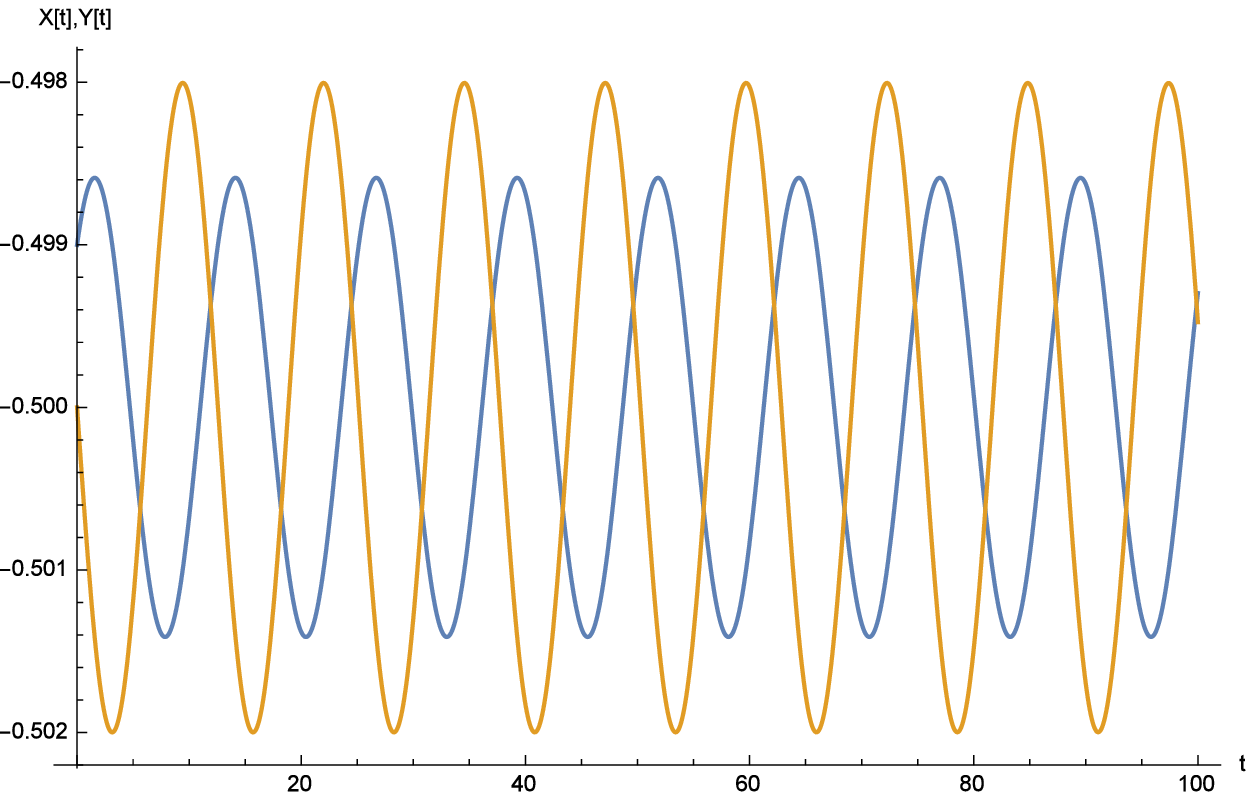}}} \resizebox*{0.4\textwidth}{!}{\rotatebox{0}
{\includegraphics{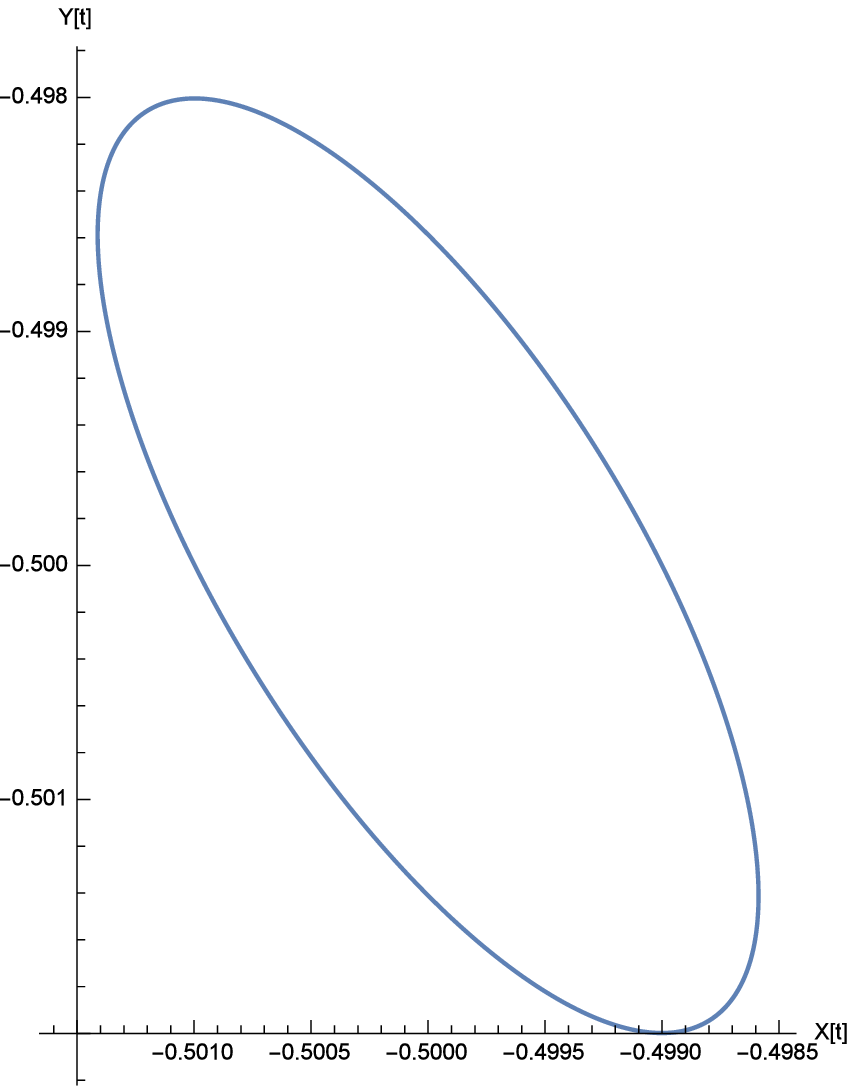}}}
 \caption{\label{figurethree} Phase plane portrait for the positive power EF Eq.~(\ref{per}) and its associated periodic solutions obtained by an Euler numerical scheme applied to the ODE system in the neighborhood of the center.}
\end{figure}

Hence the equation
\begin{equation}\label{per}
q_{_{YY}}+Y^{-5/2}q^2=0
\end{equation}
has periodic solutions, see Fig.~\ref{figurethree}.

Since now we have a center we can also write the  invariant
  \begin{equation}\label{er5}
  %........ eq 24 ..............
{\mathcal C}= Y(q_{_Y})^2-qq_{_Y}+\frac 2 3 \frac{q^3\sqrt Y}{Y^2}~.
\end{equation}
As in the previous case we are able to solve the invariant Eq.~(\ref{er5}) for the curve ${\mathcal C}=0$, to get
\begin{equation}\label{pz}
q(Y)=\frac 3 8 \sqrt Y \mathrm{sech}^2 \left(\frac{\ln Y}{4}\right)~.
\end{equation}
The above solution is not periodic. It is a particular solution valid only when ${\mathcal C}=0$, but for other values of ${\mathcal C}$, although the periodic solutions exist indeed, they can be obtained either  by numerical means in the neighborhood of the center, see Fig. \ref{figurethree}, or can be constructed analytically using the theory of elliptic equations.

 To see how analytic periodic solution are obtained for ${\mathcal C} \ne 0$, first we notice that Eq.~(\ref{per}) corresponds to Eq.~(\ref{equ1}) with $n=2$.  The solitonic solution (\ref{pz}) is obtained from system  (\ref{equ2}) choosing  $C_2=0$, and $a=C_1=1$.
Notice that for  $\lambda =\frac 1 2 $, since $\psi$ is a cubic monomial , then $\tau(\Theta)$ satisfies the elliptic equation
\begin{equation}\label{xo}
\left(\frac{d \tau}{d \Theta}\right)^2=\frac {2 \tau^3} {3}+\frac {\tau^2}{4}+C_2
\end{equation}

By choosing any  $C_2 \ne 0$, inverting (\ref{xo}), and using system (\ref{equ2}), families of smooth  and non-smooth periodic solutions with large amplitudes can also be obtained.

For this last case, its partner equation is
\begin{equation}\label{get}
\sqrt{\tilde w} \tilde w_{ss}+1=0~,
\end{equation}
which has the form discussed by Parsons for the space-charge in a plane diode \cite{par}, and later by Gettys {\em et al} \cite{gettys}.
Following the same idea of nonlinear superposition, we obtain an obvious particular solution to (\ref{get})
%........... eq 42 ..........
\begin{equation}\label{get2}
\tilde w(s)=\frac 3 2 \sqrt[3]{\frac 3 2 (1+s)^4}~.
\end{equation}

In \cite{As1}, Aslanov discusses the following generalized Emden-Fowler equation
%Indeed, according to reviewer 2, equation(3) of Ref[1] reads
\begin{equation}\label{leq1}
y\rq{}\rq{}+\frac{k+1}{x}y\rq{}+x^{2 k+kp-2}y^{2p+5}=0~,
\end{equation}
which has as particular solution \cite{As2}
%and has particular solution, see equation (15) of Ref[2]
\begin{equation}\label{leq2}
y=\left(\frac{1}{1+\frac{x^{k(p+2)}}{k^2(p+3)}}\right)^{\frac{1}{p+2}}~.
\end{equation}

In our case, we identify $k=-1,p=-\frac 32$, such that Eq.~\eqref{leq1} is Eq. \eqref{equ1} for $n=2$, or Eq.~(\ref{per}).
Thus, using \eqref{leq2}, another  particular solution to \eqref{equ1} is

\begin{equation}\label{leq3}
y=\frac{9x}{(2+3 \sqrt x)^2}~.
\end{equation}

\subsection*{6. Conclusion}
Using the phase-plane analysis of the counterpart autonomous two-dimensional ODE system  for the Emden-Fowler equations of the type (\ref{ef}),
%\begin{equation}\label{efs}
%q_{_{YY}}=\alpha Y^{-\lambda-2}q^n~,
%\end{equation}
we have proved that these equations have small amplitude smooth periodic solutions  when the trajectories lie in a neighborhood of the equilibrium $ (-1/2, 1/(2\alpha) )$ provided that the Rosenau integrability condition is satisfied, i.e., $n=2\lambda+1$, and $n>1$. This is obtained from the condition of having at least one center in the set of fixed points, which is equivalent to the nullity of the trace of the Jacobian matrix, a condition which comes out to be identical to the Rosenau first integrability condition, whereas the sufficiency implying $n>1$ is obtained from the determinant of the Jacobian matrix. Thus, we conclude that
\begin{equation}\label{efw}
q_{_{YY}}=\alpha Y^{-\lambda-2}q^{2 \lambda+1}
\end{equation}
has periodic solutions only when $\lambda>0$. These periodic solutions are of small amplitude in the immediate neighborhood of the fixed point $(-1/2, 1/(2\alpha) )$, and also of large amplitude away from the fixed point which can be found analytically by solving the elliptic equation Eq.~(\ref{xo}).

Using the  transformation $q=\frac w s$ with $Y=\frac 1 s$, any EF equation of type (\ref{efw}) becomes
\begin{equation}\label{efww}
w_{_{ss}}=\alpha s^{\lambda-1-n}q^{2 \lambda+1}.
\end{equation}
When the first Rosenau condition $n=2 \lambda+1$ is satisfied it generates the same equation
\begin{equation}\label{efwq}
w_{_{ss}}=\alpha s^{-\lambda-2}w^{2 \lambda+1}
\end{equation}
while  the partner equation obtained using first integrability condition $n=\lambda-1$ leads to  its partner equation
\begin{equation}\label{efwr}
\tilde w_{_{ss}}=\alpha  \tilde w^{\lambda-1}.
\end{equation}
Thus, one can get solutions of (\ref{efw}) from solutions of (\ref{efwq}) and vice versa, and  can generate periodic solutions of (\ref{efw}) by varying $\lambda>1$ in (\ref{efwr}).

For both integrability conditions we found the invariant (\ref{100}) when $n=2\lambda+1$ and (\ref{1005}) when $n=\lambda-1$ respectively. In  both cases, particular solutions can be calculated by varying the arbitrary constants using the systems (\ref{equ2}) and (\ref{equ5})-(\ref{equ6}), respectively.

\bigskip
\bigskip
\bigskip

{\bf Acknowledgments}\\

We wish to thank the referees for their very useful remarks.

\bigskip

{\bf References}\\

\end{document}